\documentclass{article}

\usepackage{PRIMEarxiv}

\usepackage[utf8]{inputenc} 
\usepackage[T1]{fontenc}    
\usepackage{hyperref}       
\usepackage{url}            
\usepackage{booktabs}       
\usepackage{amsfonts}       
\usepackage{nicefrac}       
\usepackage{microtype}      
\usepackage{fancyhdr}       
\usepackage{graphicx}       
\graphicspath{{media/}}     

\title{Mapping TLE orbital parameters to GNSS ephemeris for LEO PNT mega-constellation orbit simulations and visibility analysis
}

\author{
  Miquel Garcia-Fernandez \\
  Rokubun, Av. Paral·lel 88, 08015, Barcelona, Spain \\
  \texttt{miquel.garcia@rokubun.cat} \\
}

\begin{document}
\maketitle

\begin{abstract}

The emergence of Low Earth Orbit (LEO) satellite constellations dedicated to positioning applications holds the promise of improving the capabilities of existing Global Navigation Satellite Systems (GNSS). However, the absence of operational systems necessitates a qualitative assessment of potential improvements through simulation. This paper introduces a methodology to convert Two Line Element (TLE) orbital parameters—abundantly available for LEO constellations for communication and Earth Observation—into the widely used RINEX 4 format employed by GNSS.

The primary goal is to establish a comprehensive database of LEO constellation orbits directly compatible with the orbit propagation algorithms utilized in GNSS systems like the Global Positioning System (GPS). This approach enables seamless integration into simulation tools with minimal adjustments. While TLE parameters are optimized for the SGP4 propagation model and cautioned against use in classical Kepler orbit propagation scenarios requiring precision, the obtained discrepancies, within a few tens of kilometers, suggest that these representations are realistic for simulation purposes, as demonstrated with the Spire LEMUR LEO constellation.

As a practical application, the paper conducts a visibility analysis using the Starlink constellation. Results affirm expectations, showcasing that the combination of GNSS with a LEO mega-constellation significantly enhances satellite coverage and reduces Dilution of Precision. This work contributes to the ongoing discourse on the potential benefits and practicality of integrating emerging LEO constellations with established GNSS systems, offering insights into improved navigation and timing capabilities through simulation-based assessments.

\end{abstract}


\section{Introduction}
\label{sec:intro}

The emergence of Low Earth Orbit (LEO) satellite constellations dedicated to Position, Navigation, and Timing (PNT), exemplified by initiatives such as Xona or Geely (\cite{prol2022position}), signifies a transformative phase in satellite positioning technologies. While Medium Earth Orbit (MEO) Global Navigation Satellite Systems (GNSS) like GPS or Galileo have been pivotal in furnishing precise positioning, the advent of LEO PNT constellations will diversify satellite-based navigation capabilities. However, these systems are typically under the control of corporate institutions, limiting access to research data. Despite ongoing efforts in simulation studies (\cite{prol2023simulations}), the scarcity of available data hampers comprehensive performance assessments through either real or simulated datasets.

To address this challenge, this work proposes a novel approach by harnessing Two Line Element (TLE) sets, widely employed for conveying orbital information, particularly for LEO satellites. As outlined in \cite{hoots1980models}, TLE parameters are specifically tailored for orbital propagators like the Simplified General Perturbations (SGP4) model, offering LEO orbit representations with errors in the range of kilometers (\cite{sang2014achievable}).

In the domain of GNSS, the Receiver Independent Exchange Format (RINEX) stands as a universally accepted standard for representing and exchanging satellite navigation data, encompassing orbits, clocks, and raw measurements (\cite{igs2021receiver}). Designed primarily for MEO GNSS systems, RINEX ensures compatibility across diverse receiver manufacturers and analysis tools.

While simulation tools can incorporate SGP4 propagation algorithms to represent LEO PNT orbits, having this representation in a format widely adopted by the GNSS community, such as RINEX (\cite{dobbin2023flexible}), would offer convenience. This becomes particularly crucial for establishing a consistent satellite naming convention that seamlessly links satellite orbits with their corresponding raw measurement sets within the same timescale. Such standardization expedites development steps, allowing researchers/users to utilize orbit propagators described in GNSS Interface Control Documents, compatible with the orbit description in RINEX navigation files.

However, adapting LEO constellation data, especially derived from TLE parameters, to the RINEX format necessitates a specific mapping strategy, as RINEX files correspond to classical Keplerian orbital parameters. While cautionary notes (\cite{harding2019badscienceblog}) advise against this mapping for accurate orbit representation, this work acknowledges that for simulation purposes, a realistic (approximated) representation suffices. Thus, a TLE-based RINEX navigation file is proposed here, with a slight modification to RINEX 4 to accommodate the large number of LEO PNT satellites (often exceeding 100, as per \cite{garciafernandez2023leorinex}). Importantly, this strategy, mapping TLE to RINEX (TLE-RNX), circumvents the complexities of designing orbital parameters for an entire LEO mega-constellation, leveraging the abundant TLE sets available on servers like Celestrak.

This work introduces the methodology for this mapping in the first section, along with an assessment of the expected orbit accuracy achievable with the proposed method. Subsequent sections extend the application of this methodology by conducting a visibility analysis with the Starlink mega-constellation. A conclusions section wraps this paper.

\section{Methodology}
\label{sec:methodology}

This section outlines the methodology employed for the conversion of Two Line Element (TLE) orbital parameters into the RINEX format utilized in Global Navigation Satellite Systems (GNSS). As previously mentioned, this conversion aims to provide a broad representation of the orbit. The primary objective is twofold: firstly, to capitalize on the abundance of publicly available orbit files for Low Earth Orbit (LEO) constellations accessible through servers like Celestrak, and secondly, to leverage orbit propagation algorithms utilized by Medium Earth Orbit (MEO) GNSS systems, exemplified by the Global Positioning System (GPS) algorithms detailed in Table 20-IV of the GPS Interface Control Document (\cite{dod2022gpsicd}).

By doing this, simulation tools could benefit from orbit representation in currently existing formats (with minimum modifications, as shown in e.g. \cite{garciafernandez2023leorinex}). 

The mapping strategy used is described in the following steps, based on the necessary parameters of the RINEX file (and its computation using the related TLE parameter)

\begin{itemize}
    \item Time of ephemeris ($toe$) is the epoch specified in the TLE field (fields 7 and 8 from line 1) assuming UTC timescale.
    \item Mean Anomaly at Reference Time ($M_0$), is also the mean anomaly of the TLE (field 7 from line 2) converted to radians
    \item Eccentricity ($e$), is also the eccentricity of the TLE (field 5 of line 2)
    \item Square Root of the Semi-Major Axis ($\sqrt{A}$) can be computed from the mean motion of the TLE ($n$, field 8 of line 2) and the Earth Gravitation constant ($\mu = 3.986005e14 m^3/s^2$), using the expression:
    \begin{equation}
        \sqrt{A} = \frac{\sqrt[6]{\mu}}{\sqrt[3]{n}}
    \end{equation}
    \item Longitude of Ascending Node of Orbit Plane at Weekly Epoch ($\Omega_0$) is the parameter that needs further attention for two reasons: it needs to be provided in an Earth Centered Earth Fixed reference frame (the TLE parameters are referred to an Earth Centered Inertial reference frame) and it needs to be referred to the start of the week (not at the reference epoch $toe$), in order to be compatible with the computation of the "\emph{Corrected longitude of ascending node}" ($\Omega_k$) specified in Table 20-IV of the GPS Interface Control Document (\cite{dod2022gpsicd}), reproduced here for convenience ($\dot{\Omega}_e$ is the Earth rotation rate):
    
    \begin{equation}
    \Omega_k = \Omega_0 + ( \dot{\Omega} - \dot{\Omega}_e ) \cdot tk - \dot{\Omega}_e \cdot toe
    \end{equation}

    Therefore, to compute $\Omega_0$, the following steps are necessary:
    \begin{enumerate}
        \item Take the TLE ephemeris epoch and compute the weeks and seconds within the week elapsed since an arbitrary epoch ($t_0)$. The choice of $t_0$ is not critical as long as it starts on Sunday (due to the specification in e.g. GPS system). Therefore, a possible date (actually used in this work) is the GPS reference epoch ($t_0$ set to January 6th 1980). The epoch corresponding to the start of the week ($t_{week}$) is the epoch reconstructed from the number of weeks since $t_0$ (i.e. ignoring the seconds of the week).
        \item Compute the Right Ascension of the Greenwich meridian at $t_{week}$, i.e. $\Omega_{G}(t_{week})$. This can be computed with the local sidereal time at longitude $0^\circ$ and $t_{week}$ and then transforming from hours to degrees/radians. Open source tools such as Poliastro \cite{rodriguez2016poliastro} (using the method \texttt{poliastro.earth.util.get\_local\_sidereal\_time}) can be used for this purpose.
        \item Subtract the right ascension of the ascending node of the TLE ($\Omega_{TLE}$, field 4 of line 2) so that: 
            
        \begin{equation}
        \Omega_0 = \Omega_{TLE} - \Omega_{G}(t_{week})
        \end{equation}

        You might also need to compute the modulus operation to ensure that $\Omega_0 \in [0, 2\pi)$.
    \end{enumerate}

    \item Inclination Angle at Reference Time ($i_0$) is the inclination of the TLE (field 3, line 2), converted to radians.
    \item Argument of Perigee ($\omega$) is the argument of perigee of the TLE (field 6, line 2), converted to radians.
    \item Mean Motion Difference From Computed Value ($\Delta n$), which will be later multiplied by the time elapsed since the $toe$ corresponds to the first derivative of the mean motion (i.e. ballistic coefficient, field 9 of line 1).
\end{itemize}

Other parameters such as rate of Right Ascension ($\dot \Omega$), rate of inclination angle ($idot$) and the amplitude of sine and cosine harmonics ($C*$) are set to 0.

\subsection{Accuracy of TLE-RNX}

\begin{figure}
  \centering
  \includegraphics{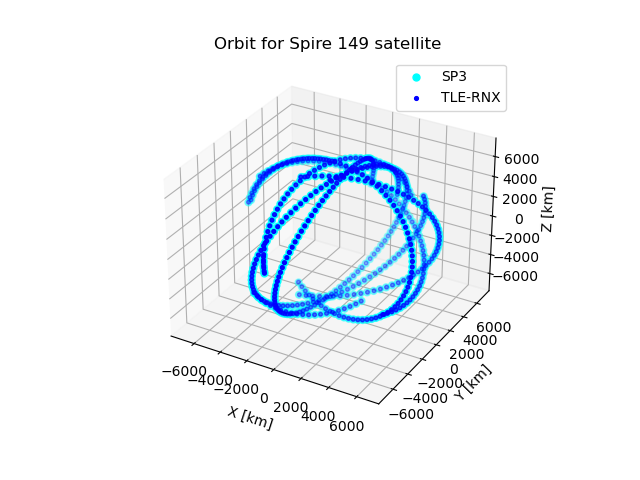}
  \caption{Sample orbit for Spire satellite 149 obtained from SP3 and orbit propagation using TLE-RNX for July 4th 2023.}
  \label{fig:tlernx_vs_sp3_spire_orbit}
\end{figure}

As previously articulated, the purpose of generating TLE-RNX files (i.e., RINEX navigation files, as delineated in \cite{igs2021receiver} and \cite{garciafernandez2023leorinex}) for Low Earth Orbit (LEO) constellations based on Two Line Element (TLE) parameters is not centered on furnishing highly precise or accurate orbital information. Instead, the aim is to establish a comprehensive database comprising orbital parameters that align with the requisites of orbital propagators defined in the Global Navigation Satellite Systems (GNSS) Interface Control Documents. Notably, the TLE parameters are specifically tailored for compatibility with the SGP4 orbit propagation model (see \cite{hoots1980models}), and are not intended for interpretation as classical Keplerian orbital parameters. Despite this warning notice, they have been utilized in this study as a coarse representation of the satellite orbits (a sample orbit is shown in Figure \ref{fig:tlernx_vs_sp3_spire_orbit}).

\begin{figure}
  \centering
  \includegraphics{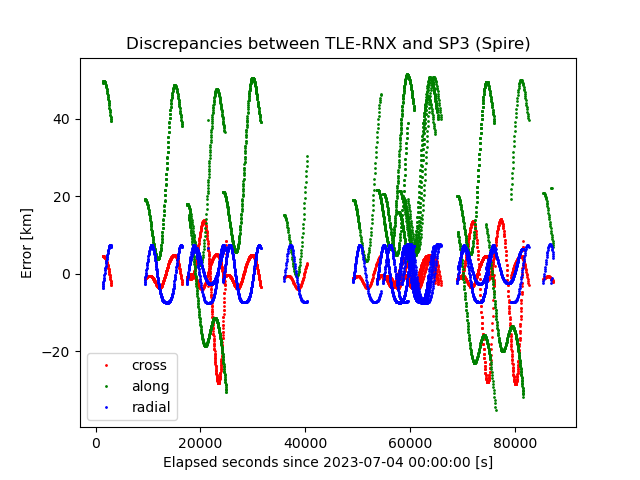}
  \caption{Single day comparison of Spire satellite positions from SP3 files and positions computed from orbital parameters derived from TLE files.}
  \label{fig:tlernx_vs_sp3_spire}
\end{figure}

To assess the disparities achievable through the proposed approach in constructing a massive database of Low Earth Orbit (LEO) satellite orbits in RINEX format, the Spire constellation serves as a pertinent case study. The rationale behind selecting Spire satellites lies in the fact that, on the one hand, the TLE data for these satellites are included in the Celestrak database, and on the other hand, precise orbit representation in SP3 format is also available in the Radio-Occultation COSMIC UCAR product database (i.e.  \texttt{leoOrb} product). The discrepancies between the two orbits (TLE-RNX and SP3) are shown in Figure \ref{fig:tlernx_vs_sp3_spire} for a single day (July 4th 2023) and for the cross-, along- and radial components. 

For the radial and cross components, maximum discrepancies of approximately 10 km can be anticipated, while larger discrepancies, reaching up to 50 km, are evident in the along component. Notably, these discernible discrepancies exhibit a well-defined pattern and repeatability, underscoring the inadequacy of the proposed approach as a precise representation of the orbit. Nevertheless, these discrepancies, although non-negligible, remain sufficiently modest to be considered valuable for simulation purposes. This holds true, at least until operational Low Earth Orbit (LEO) Positioning, Navigation, and Timing (PNT) constellations are deployed, and their operational institutions release the accurate orbital parameters.

\section{Visibility analysis}

A practical application of the proposed strategy for generating a database of Low Earth Orbit (LEO) orbits in the standard RINEX format involves conducting a comprehensive satellite visibility analysis for specific locations. This analysis helps to evaluate the enhancements attainable through the utilization of LEO constellations in combination with GNSS concerning Dilution of Precision (DOP) and general coverage. As it is already known, for MEO GNSS, there exists a satellite deficit in the North azimuths in northern mid-latitudes, and visibility is notably compromised in high latitudes, where no satellites are observable in the zenith direction (high elevation). 

\begin{figure}[h]
  \centering
  \begin{tabular}{cc}
    \includegraphics[width=0.45\textwidth]{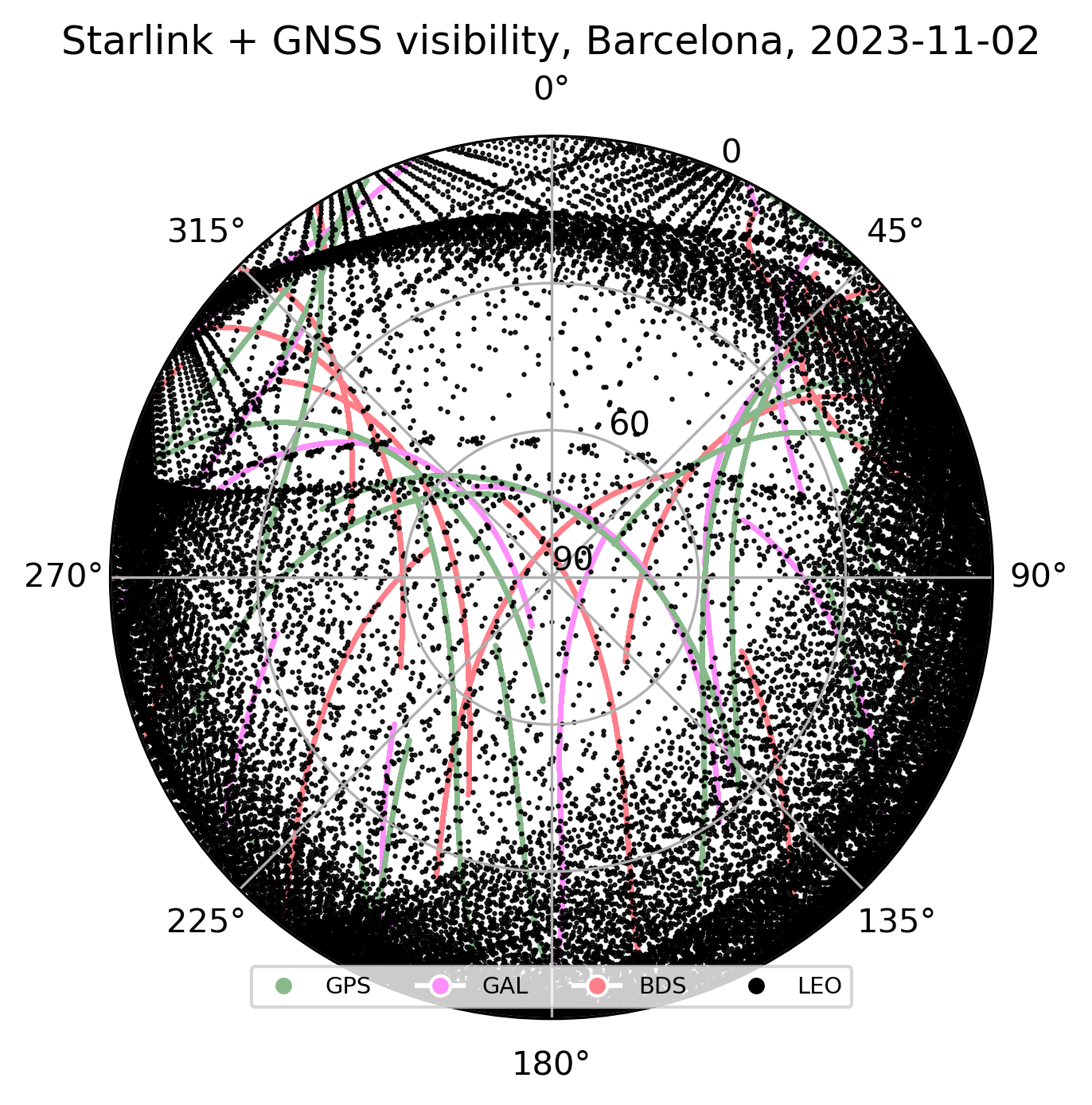} &
    \includegraphics[width=0.45\textwidth]{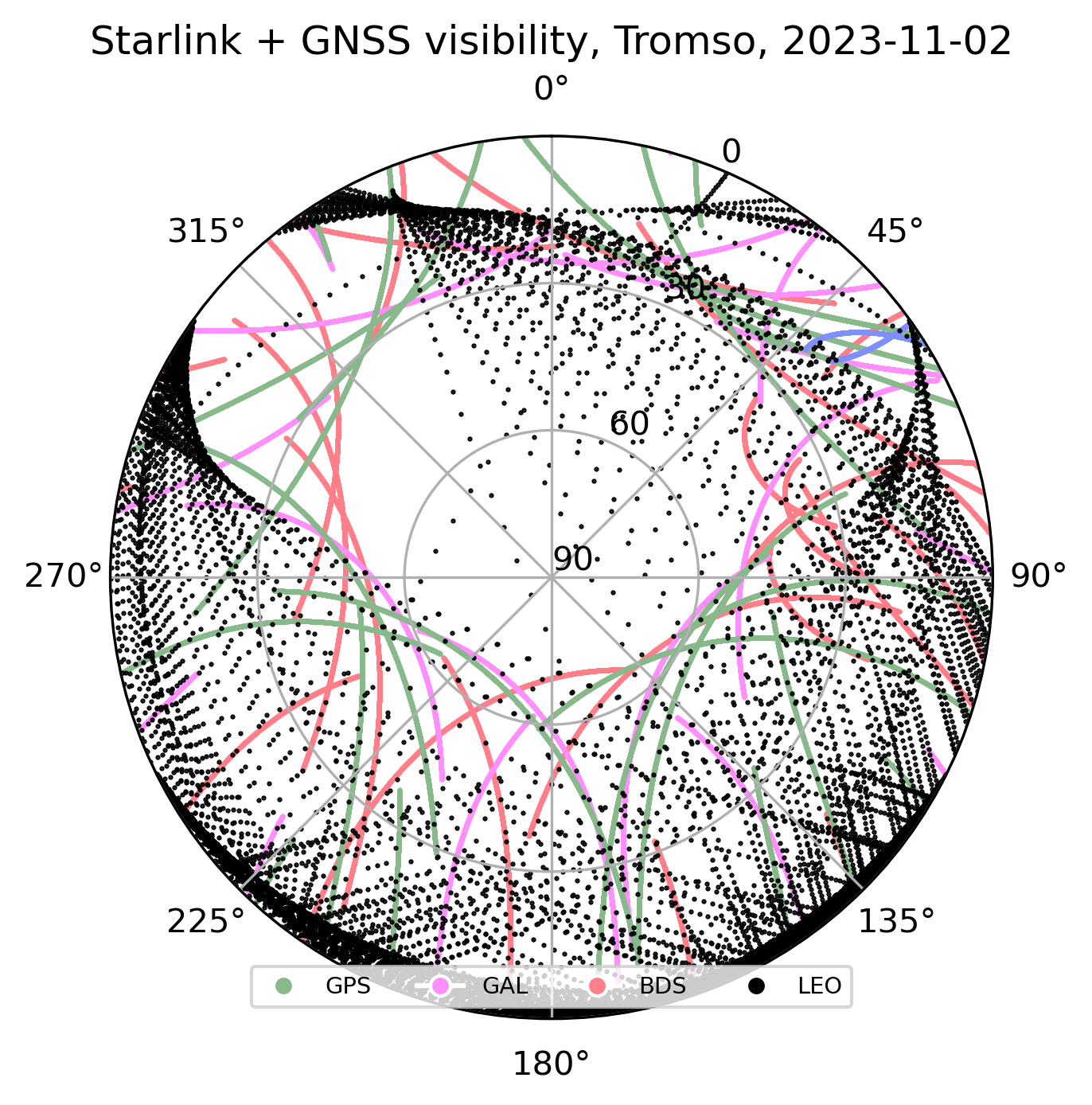}
  \end{tabular}
  \caption{Starlink and GNSS combined visibility on 2023-11-02 (between 12h and 16h) for two locations (left panel) Barcelona, Spain (mid-latitude) and (right panel) Tromsø, Norway (high-latitude}
  \label{fig:skyplot}
\end{figure}

\begin{figure}[h]
  \centering
  \begin{tabular}{cc}
    \includegraphics[width=0.45\textwidth]{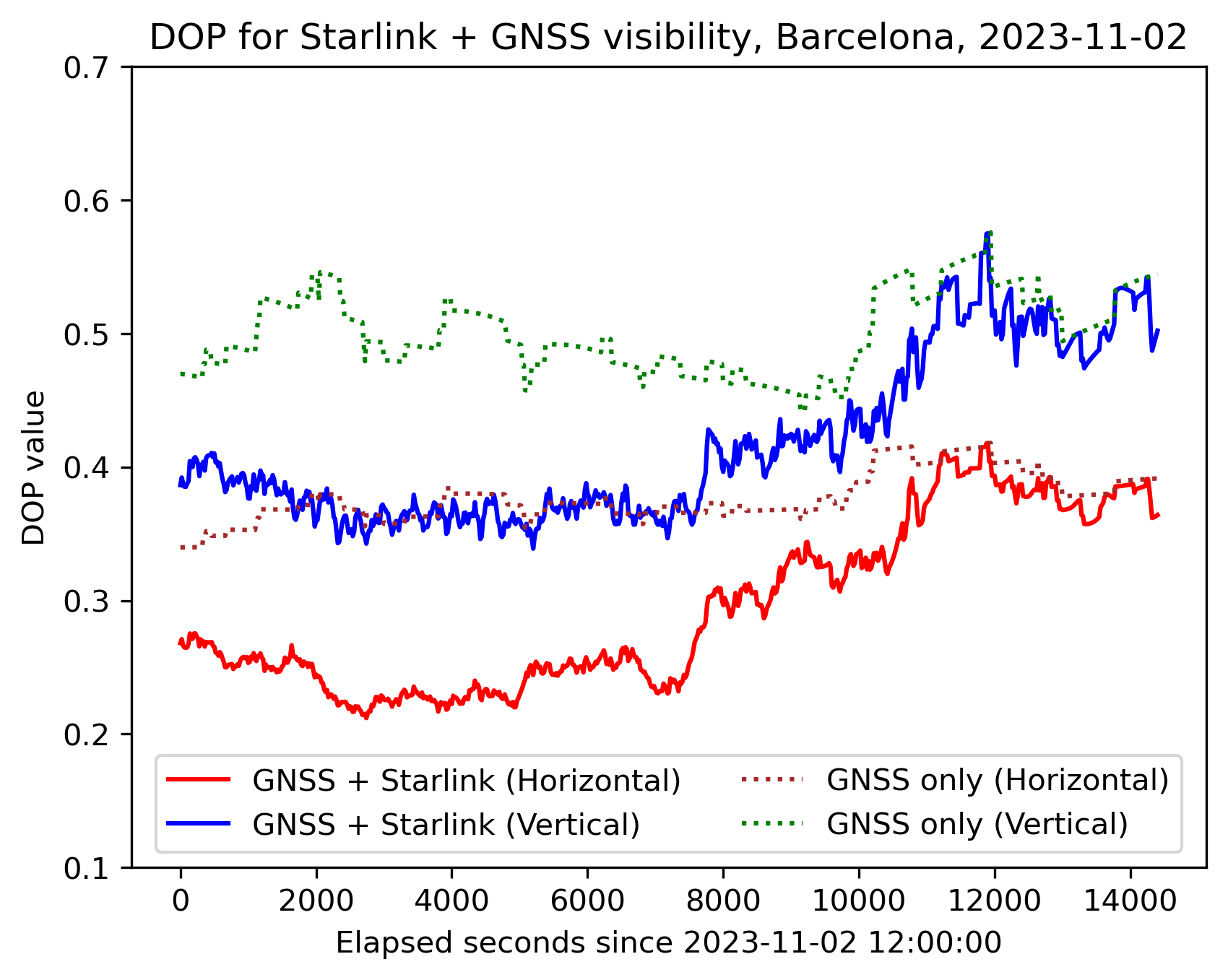} &
    \includegraphics[width=0.45\textwidth]{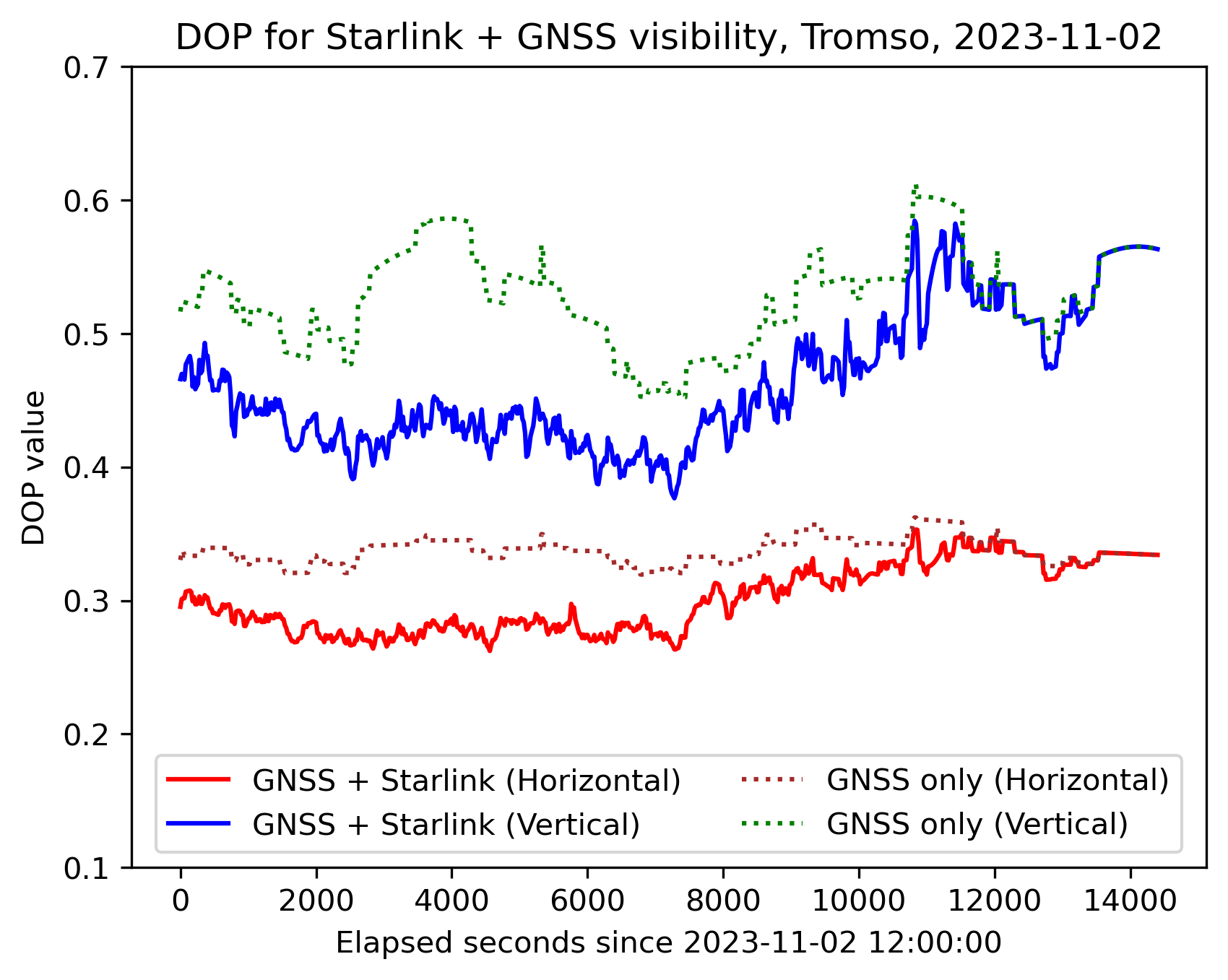}
  \end{tabular}
  \caption{Dilution of precision for GNSS and GNSS + Starlink on 2023-11-02 (between 12h and 16h) for two locations (left panel) Barcelona, Spain (mid-latitude) and (right panel) Tromsø, Norway (high-latitude}
  \label{fig:dop}
\end{figure}

As shown in Figure \ref{fig:skyplot}, augmenting MEO GNSS constellations with LEO ones, specially if highly populated with a mega-constellation such as Starlink, improves the satellite visibility and coverage. This is specially true in regions where MEO GNSS constellations may fall short (notably, addressing gaps in the north for locations at northern mid-latitudes and enhancing visibility at high elevations in high-latitude areas). This improvement is also shown in Figure \ref{fig:dop}, where the Dilution of Precision (DOP) is shown for two locations and for the cases with only GNSS as well as the combined GNSS + LEO (Starlink) case. 

\section{Conclusion}

This paper introduces a methodology for mapping Two Line Element (TLE) parameters of Low Earth Orbit (LEO) mega-constellations into orbital parameters compatible with the RINEX 4 format, enabling seamless integration with Global Navigation Satellite Systems (GNSS) orbit propagation algorithms. As it is known, the TLE parameters are finely tuned for the SGP4 propagation model and not suited for classical Keplerian orbit propagation models. However, the discrepancies observed with the proposed approach relative to precise SP3 orbits, for the Spire LEMUR LEO constellation, which amounts to tens of km, are within reasonable limits to use this strategy as a tool for simulation.

The visibility analysis using TLE-based RINEX files for the Starlink mega-constellation, boasting approximately 5000 LEO satellites, affirms the expected advantages of combining Medium Earth Orbit (MEO) and LEO GNSS. The results showcase a substantial enhancement in satellite coverage and a reduction in Dilution of Precision values, reinforcing the assumption that LEO Positioning, Navigation, and Timing (PNT) systems hold potential for improving position robustness, particularly in terms of convergence time.

The wealth of TLE data available for LEO constellations in communication and Earth Observation sets the stage for creating a comprehensive database of orbits in a slightly modified RINEX 4 format, so that these simulated LEO can be used for PNT purposes. This lays the foundation for qualitative assessments through simulation, addressing the current lack of operational systems. In this context, our work serves as a stepping stone for further advancements in simulation steps, including the computation of synthetic raw measurements (pseudorange, carrier phase, Doppler) from LEO PNT constellations. Incorporating error factors such as ionosphere and troposphere, and advancing hybridization algorithms between MEO and LEO GNSS, will further enhance key performance indicators such as accuracy, coverage, convergence time, and integrity.

\section*{Acknowledgments}

The author extend their gratitude to the team at the Earth Intelligence Department of Spire Global UK for sharing insights into the correlation between NORAD ID catalog numbers and satellite identification numbers as specified in the SP3 files hosted at the UCAR Radio Occultation server. Their contribution has significantly helped to obtain the results presented in this paper.

The author wish to thank also Celestrack and COSMIC UCAR to make available the TLE element sets and SP3 orbit files respectively. This data is available at their servers \url{https://celestrak.org/} and  (\url{https://data.cosmic.ucar.edu/gnss-ro}). 

\bibliographystyle{unsrt}  
\bibliography{main}

\end{document}